\documentclass[traditabstract,article]{aa}
\pdfoutput=1
\usepackage{txfonts}
\usepackage{natbib}
\usepackage{graphicx}
\usepackage{afterpage}
\usepackage{lscape}
\usepackage{psfrag}
\usepackage{url}
\usepackage{microtype}
\usepackage{float}
\usepackage{afterpage}
\usepackage{longtable}
\usepackage{fixltx2e}
\usepackage[colorlinks=true, linkcolor=blue, citecolor=blue, urlcolor=blue]{hyperref}
\usepackage{breakurl}
\usepackage{xcolor}
\usepackage{flushend}
\usepackage[bottom]{footmisc}

\def\apjl{ApJL}

\def\aap{A\&A}
\def\apjs{ApJS}

\makeatletter
\renewcommand*{\@fnsymbol}[1]{$\star$}
\makeatother

\begin{document}
   \title{The spatially-resolved correlation between [N{\sc ii}]~205~$\mu$m line emission and the 24~$\mu$m continuum in nearby galaxies}

   \subtitle{ }

   \author{T. M. Hughes\inst{1,2,}\thanks{email: thomas.hughes@uv.cl}, M. Baes\inst{2}, M. R. P. Schirm\inst{3}, T. J. Parkin\inst{3}, R. Wu\inst{4}, I. De Looze\inst{2}, C. D. Wilson\inst{3}, \\S. Viaene\inst{2}, G. J. Bendo\inst{5}, A. Boselli\inst{6}, D. Cormier\inst{7}, E. Ibar\inst{1}, O. \L. Karczewski\inst{8}, N. Lu\inst{9}, L. Spinoglio\inst{10}}
	  
\institute{Instituto de F\'{i}sica y Astronom\'{i}a, Universidad de Valpara\'{i}so, Avda. Gran Breta\~{n}a 1111, Valpara\'{i}so, Chile
\and  Sterrenkundig Observatorium, Universiteit Gent, Krijgslaan 281-S9, Gent 9000, Belgium
\and Department of Physics \& Astronomy, McMaster University, Hamilton, Ontario L8S 4M1, Canada 
\and Department of Astronomy, the University of Tokyo, Bunkyo-ku, Tokyo 113-0033, Japan 
\and UK ALMA Regional Centre Node, Jodrell Bank Centre for Astrophysics, School of Physics and Astronomy, \\ University of Manchester, Oxford Road, Manchester M13 9PL, UK
\and Laboratoire d'Astrophysique de Marseille, Université d'Aix-Marseille and CNRS, UMR7326, F-13388 Marseille Cedex 13, France   
\and Institut für theoretische Astrophysik, Zentrum für Astronomie der Universität Heidelberg, Albert-Ueberle 2, Heidelberg, Germany
\and Department of Physics \& Astronomy, University of Sussex, Brighton, BN1 9QH, UK 
\and Infrared Processing and Analysis Center, California Institute of Technology, MS 100-22, Pasadena, CA 91125, USA
\and Istituto di Astrofisica e Planetologia Spaziali, INAF-IAPS, Via Fosso del Cavaliere 100, I-00133 Roma, Italy } 

   \date{Accepted for publication in A\&A.}

\newcommand{\hi}{H{\sc i}} 
\newcommand{\hii}{H{\sc ii}\ }
\newcommand{\cii}{[C{\sc ii}]}
\newcommand{\ci}{[C{\sc i}]}
\newcommand{\oi}{[O{\sc i}]}
\newcommand{\oii}{[O{\sc ii}]}
\newcommand{\oiii}{[O{\sc iii}]}
\newcommand{\oiv}{[O{\sc iv}]}
\newcommand{\nii}{[N{\sc ii}]}
\newcommand{\niii}{[N{\sc iii}]}
\newcommand{\ha}{H{\sc $\alpha$}}
\newcommand{\hd}{H{\sc $\delta$}}
\newcommand{\hg}{H{\sc $\gamma$}}
\newcommand{\hb}{H{\sc $\beta$}}
\newcommand{\kms}{km~s$^{-1}$\ }
\newcommand{\sdust}{$\Sigma_{\mathrm{dust}}$}
\newcommand{\sgas}{$\Sigma_{\mathrm{gas}}$}
\newcommand{\shii}{$\Sigma_{\mathrm{H}_{2}}$} 
\newcommand{\shi}{$\Sigma_{\mathrm{H}\tiny{\textsc{i}}}$} 
\newcommand{\ssfr}{$\Sigma_{\mathrm{SFR}}$}

  \abstract{A correlation between the 24~$\mu$m continuum and the \nii \ 205~$\mu$m line emission may arise if both quantities trace the star formation activity on spatially-resolved scales within a galaxy, yet has so far only been observed in the nearby edge-on spiral galaxy NGC 891. We therefore assess whether the \nii \ 205 -- 24~$\mu$m emission correlation has some physical origin or is merely an artefact of line-of-sight projection effects in an edge-on disc. We search for the presence of a correlation in \textit{Herschel} and \textit{Spitzer} observations of two nearby face-on galaxies, M51 and M83, and the interacting Antennae galaxies NGC 4038 and 4039. We show that not only is this empirical relationship also observed in face-on galaxies, but also that the correlation appears to be governed by the star formation rate (SFR). Both the nuclear starburst in M83 and the merger-induced star formation in NGC 4038/9 exhibit less \nii \ emission per unit SFR surface density than the normal star-forming discs. These regions of intense star formation exhibit stronger ionization parameters, as traced by the 70/160 $\mu$m far-infrared colour, that suggest the presence of higher ionization lines that may become more important for gas cooling, thereby reducing the observed \nii ~205~$\mu$m line emission in regions with higher star formation rates. Finally, we present a general relation between the \nii ~205~$\mu$m line flux density and SFR density for normal star-forming galaxies, yet note that future studies should extend this analysis by including observations with wider spatial coverage for a larger sample of galaxies.}

     \keywords{galaxies: star formation --
             galaxies: spiral --
            galaxies: ISM --
            infrared: galaxies --
             ISM: lines and bands
               } 

	\authorrunning{T. M. Hughes et al.}
	\titlerunning{\nii \ 205~$\mu$m line vs. 24~$\mu$m continuum correlation}
   \maketitle

\section{Introduction}

Diagnostic tracers of the star formation rate (SFR) have traditionally been based on observations of the broad band continuum (e.g. FUV, 24~$\mu$m band) or spectral line emission (e.g. H$\alpha$, P$\alpha$) of the Milky Way and nearby galaxies (see \citealp*{kennicutt2012} for a review, and references therein). The recent success of the {\it Herschel Space Observatory} \citep{pilbratt2010} is enabling the development of new SFR indicators at far-infrared (FIR) wavelengths (e.g. \citealp{li2013}), particularly with the FIR fine-structure lines (e.g. \citealp{delooze2011}; \citealp{farrah2013}; \citealp{sargsyan2014}; \citealp{delooze2014}; \citealp{pineda2014}; \citealp{herreracamus2014}). In the future, ground-based interferometers capable of observing these lines on spatially-resolved (sub-arcsecond) scales, such as the Atacama Large Millimetre/sub-millimetre Array (ALMA) and the Northern Extended Millimetre Array (NOEMA), will allow us to trace the star formation activity out to the high-redshift Universe. 

The \nii \ 205~$\mu$m line is one such line, which originates from the $^3$P$_1$ $\rightarrow$ $^3$P$_0$ transition of the ground state of singly ionized nitrogen. Nitrogen's ionization potential of 14.53 eV, slightly larger than that of hydrogen, means the \nii \ emission traces all of the warm ionized interstellar medium (ISM) and, with a critical density for collisions with electrons of only 44~cm$^{-3}$ at $T=8000$~K \citep{oberst2006,oberst2011}, it is highly susceptible to collisional excitation. The \nii \ line also has the advantage of typically being optically thin due to a small Einstein coefficient and, like other FIR lines, does not suffer from the problem of obscuration by interstellar dust that affects UV/optical tracers. Therefore, \nii \ is expected to be an excellent indicator of SFR via measurement of the flux of ionizing photons.   

For a sample of 70 galaxies in the \textit{Herschel} Spectroscopic Survey of Warm Molecular Gas in Local Luminous Infrared Galaxies, \citet{zhao2013} found that the SFR determined from the total infrared luminosity via the relationship in \citet*{kennicutt2012} correlates with the integrated \nii \ 205~$\mu$m line luminosity. More recently, \citet{wu2014} found a spatially-resolved correlation between the surface densities of the SFR and \nii \ 205~$\mu$m line in the M83 galaxy, and an intersection of the local relationship and the global relationship of \citet{zhao2013} at high $\Sigma_{\mathrm{SFR}}$ that suggests the latter correlation is dominated by active star-forming regions. Since the 24~$\mu$m and the \nii \ 205~$\mu$m line emission both seem to trace the SFR on spatially-resolved scales, one might expect to find a relationship between these two quantities.

In a recent study of the nearby edge-on spiral galaxy NGC~891, \citet{hughes2014a} exploited such an empirical relationship to predict the \nii \ 205~$\mu$m line emission from a higher resolution 24~$\mu$m image. Although this appears to be a powerful and promising technique for estimating the \nii \ 205~$\mu$m line emission from more widely available \textit{Spitzer}-MIPS data, the underlying nature of the correlation is still unknown due in part to uncertainties associated with observations of high inclination systems. Integration along the line-of-sight through an edge-on spiral galaxy will include a range of physical environments, from active star-forming \hii \ regions to diffuse ionised gas, making it difficult to interpret the origin of the relationship. In this paper, we thus attempt to assess whether the \nii \ 205 -- 24~$\mu$m correlation has some physical origin or is merely an artefact of line-of-sight projection effects in an edge-on disc. We search for the presence of a correlation in \textit{Herschel} and \textit{Spitzer} observations of face-on and interacting galaxies, and compare our findings to the NGC~891 correlation of \citet{hughes2014a}. We show that not only is this empirical relationship also observed in the face-on systems, but also that correlation appears to be driven by the star formation rate. In the next section, we describe the data used for our analysis and in Sec. 3 we present our results. In Sec. 4 and 5, we discuss our results and conclusions, respectively.

\begin{figure*}
\begin{center}
\includegraphics[width=0.95\textwidth]{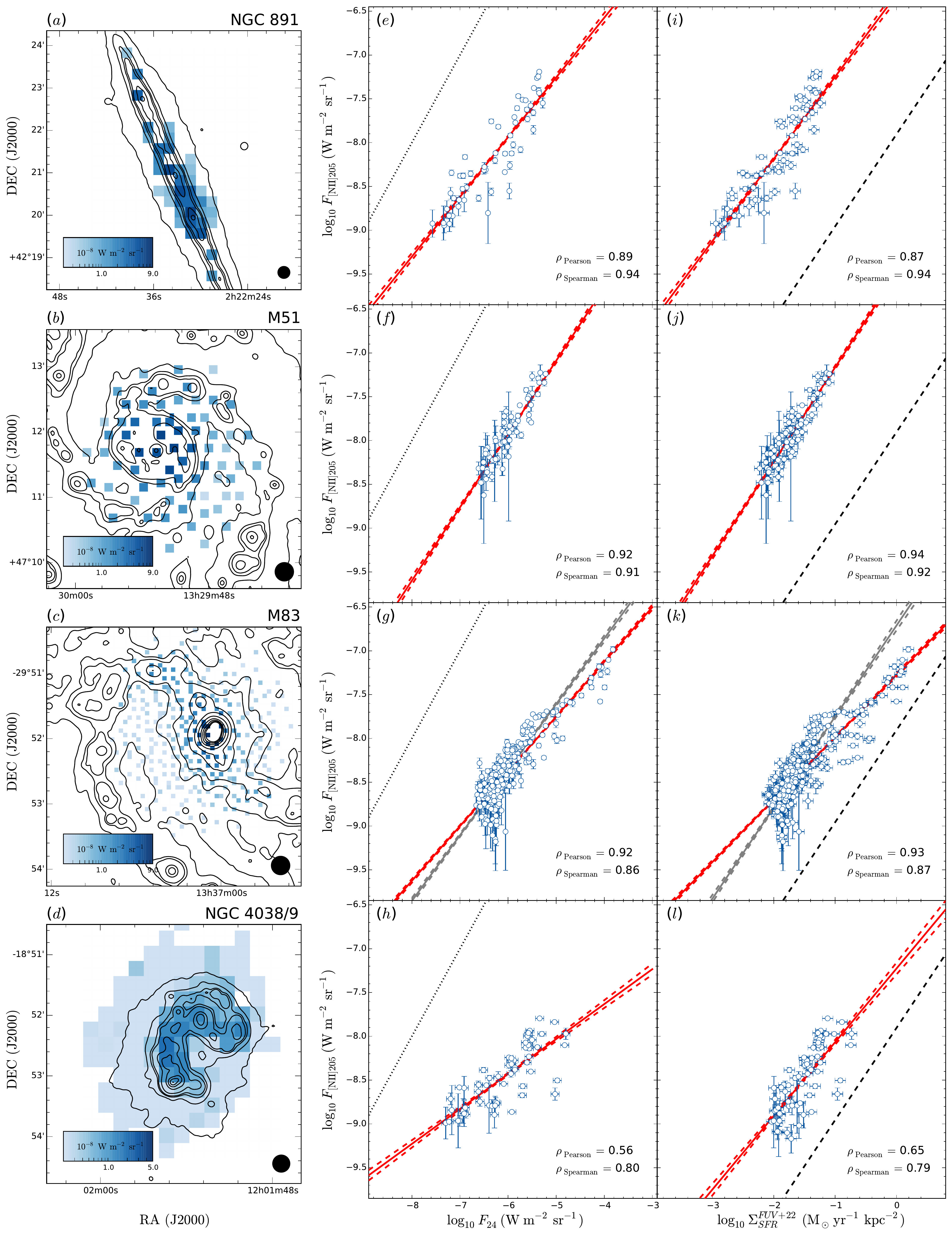}
\end{center}
\vspace{-0.4cm}
\caption{For each galaxy (\textit{rows}), we present the SPIRE FTS \nii ~205~$\mu$m line emission maps at 17$\arcsec$ resolution (i.e. the beam size, indicated by a circle) superimposed on contours of constant 24~$\mu$m emission (\textit{left panels}). The observed logarithms of the \nii \ 205 $\mu$m line flux density versus 24~$\mu$m continuum flux density (\textit{middle panels}) and $\Sigma_{SFR}^{FUV+22}$ (\textit{right panels}) are shown for each resolution element (blue circles). The scatter plots include the best linear fit (red solid line), 1$\sigma$ confidence intervals (red dashed lines), the 1:1 relationship (dotted line), and the converted \nii \ 205-SFR relation from integrated measurements (black dashed line) found by \citet{zhao2013}. The grey lines for M83 indicate the best fit and confidence intervals obtained when excluding central pixels ($\log_{10} F_{24} (\mathrm{W~m}^{-2}~\mathrm{sr}^{-1}) > -4.4$). We state the Pearson and Spearman coefficients.}\label{fig:correlation}
\end{figure*}

\section{The data and image processing}

The Very Nearby Galaxy Survey (VNGS; P.~I.:~C.~D.~Wilson; see e.g. \citealp{parkin2013,parkin2014}; \citealp{schirm2014}; \citealp{hughes2014b}), a \textit{Herschel} Guaranteed Time Key Project that aims to study the gas and dust in the ISM of a diverse sample of 13 nearby galaxies using \textit{Herschel}'s PACS and SPIRE instruments (see \citealp{pilbratt2010}; \citealp{poglitsch2010}; \citealp{griffin2010}), provides a unique dataset for investigating the spatially-resolved \nii ~205~$\mu$m line emission. Observations with the SPIRE Fourier Transform Spectrometer (FTS) of the \nii \ 205~$\mu$m line reach a $\sim$17$\arcsec$ spatial resolution. Because we require enough spatial coverage to accurately define and study the correlations, we select the late-type galaxies that have been observed in the high spectral resolution, intermediate (rather than sparse) spatial sampling mode. Our selection criteria include five objects: the edge-on spiral NGC 891, face-on spirals M51 and M83, and the interacting Antennae galaxies NGC~4038 and 4039 (4038/9 hereafter). Just as NGC 891 is the member of the sample that typifies an edge-on spiral, M51 and M83 are representative of face-on spiral galaxies and thus the most logical choices for a comparison with NGC~891. We also include the merging galaxies NGC~4038/9 to draw comparisons with more intense star-forming regions. Below, we summarise the available data and image processing steps.

\subsection{SPIRE FTS \nii \ line observations}

The SPIRE FTS instrument \citep{griffin2010} consists of two bolometer arrays, the SPIRE Short Wavelength (SSW) array and the SPIRE Long Wavelength (SLW) array, covering wavelength ranges of 194 to 313 $\mu$m and 303 to 671 $\mu$m, respectively, with a 2$\arcmin$ diameter field of view. A complete description of the instrument may be found in the SPIRE Observers' Manual\footnote{Document HERSCHEL-DOC-0798 version 2.5 (March 2014), is available from the ESA \emph{Herschel} Science Centre at \burl{http://herschel.esac.esa.int/Docs/SPIRE/spire_handbook.pdf}}. Intensity maps of the \nii ~205~$\mu$m line emission for NGC 891, M51, M83 and NGC~4038/9 have been presented in \citealp{hughes2014a}, \citealp{parkin2013}, \citealp{wu2014} and \citealp{schirm2014}, and we adopt the same distances of 9.6, 8.4, 3.6, and 20.0~Mpc, respectively. Here, we briefly summarise the observations and data reduction for each galaxy, whilst full details of may be found in these corresponding papers.  

In general, observations are reduced using HIPE with a SPIRE calibration context later than v.11.0.1 and basic standard processing steps. The observations for each galaxy consist of one central pointing in high spectral resolution intermediate-sampling mode, whilst for NGC~891 we also include publicly available open time observations (P.~I.:~G.~Stacey) taken in sparse-sampling mode of the upper and lower portions of the disc. In all cases, we use the point-source calibrated data. Two spectral cubes corresponding to the SLW and SSW arrays are created via the \emph{spireProjection} task, and the final spectra are fit with a polynomial and Sinc function for the baseline continuum and line emission. We note that \citet{wu2014} use the MPFIT procedure \citep{markwardt2009} to fit the continuum and \nii \ 205~$\mu$m line with parabolic and Sinc functions, before the continuum-subtracted spectra are co-added in each area. Since all but M83 were previously reduced using a homogeneous methodology, described in detail in \citet{schirm2014}, we chose to reprocess the M83 observations in the same manner, using HIPE v.12.1 and SPIRE calibration context v.12.3. The final integrated \nii \ flux maps have a 15$\arcsec$ pixel scale in the cases of NGC~891 and NGC~4038/9, whereas those of M51 and M83 have a $4\arcsec$ pixel scale such that the finite pixels are centred on each of the bolometers of the FTS array, while the remaining pixels are left blank (c.f. the \nii \ maps in \hyperref[fig:correlation]{Fig.~\ref*{fig:correlation}}, left column). Our reprocessed M83 intensity map is consistent within the errors to that of \citet{wu2014}.

\subsection{MIPS 24~$\mu$m continuum observations}

Our five chosen galaxies have 24~$\mu$m data obtained with the Multiband Imaging Photometer for {\it Spitzer} (MIPS; \citealp{rieke2004}) and reprocessed by \citet{bendo2012} using the MIPS Data Analysis Tools \citep{gordon2005} with some additional processing. Each image has a pixel scale of $1\farcs5$, a PSF FWHM of $6\arcsec$, and calibration uncertainties of $\sim$4\% \citep{engelbracht2007}. To facilitate a comparison between the \nii \ intensity maps, our 24~$\mu$m maps were first convolved to the 17$\arcsec$ resolution of the \nii ~205~$\mu$m image using the common-resolution convolution kernels of \citet{aniano2011}, and rescaled to the appropriate pixel scale for each galaxy. Although these pixel sizes are typically smaller than the beam size, i.e. adjacent pixels are not independent, we note that the analysis reproduces the same trends and conclusions as found when setting the pixel size equal to the beam size in each case. At the adopted distances, physical scales are 0.70, 0.61, 0.34 and 1.45~kpc for NGC 891, M51, M83 and NGC~4038/9, respectively.

\section{Results}

\subsection{\nii \ 205~$\mu$m line vs. 24~$\mu$m continuum correlation}

Our first goal is to confirm the existence of a correlation in \textit{Herschel} and \textit{Spitzer} observations of M51 and M83, and compare our findings to those of NGC~891. Such a correlation is already promisingly evident in the maps presented in the left column of \hyperref[fig:correlation]{Fig.~\ref*{fig:correlation}}: qualitatively, the brighter regions of the \nii \ emission appear to correspond to the brighter contours of the 24~$\mu$m emission. In the middle column of \hyperref[fig:correlation]{Fig.~\ref*{fig:correlation}}, we quantify the correlation between \nii ~205~$\mu$m line emission and MIPS 24~$\mu$m emission for each galaxy. \hyperref[fig:correlation]{Fig.~\ref*{fig:correlation}e} presents the original correlation in NGC~891 reported by \citet{hughes2014a}. To briefly recap their results, the Spearman ($\rho_\mathrm{S}$) and Pearson ($\rho_\mathrm{P}$) coefficients of rank correlation are 0.94 and 0.88, respectively, where a value of 1 represents a perfect correlation. The best fit linear relation is
\begin{eqnarray}\label{eqn:n891corr}
\log_{10}\, F_{\mathrm{[NII]}205} = (0.77\pm 0.01)\,\log_{10}\, F_{24} - (3.31\pm 0.04)
\end{eqnarray}
\noindent where both flux densities are in units of W m$^{-2}$ sr$^{-1}$. In \hyperref[fig:correlation]{Fig.~\ref*{fig:correlation}f} and \hyperref[fig:correlation]{\ref*{fig:correlation}g}, we present our results for the face-on galaxies, M51 and M83. The former galaxy shows a correlation of striking similarity to that of NGC~891. In fact, the Spearman and Pearson coefficients of 0.91 and 0.92 indicate a strong, linear correlation in log-log space. A best linear fit, expressed as Eq.~\ref{eqn:n891corr} above, yields a gradient of 0.83$\pm$0.02 and intercept equal to -2.90$\pm$0.09, i.e. very similar to the best fit relation to the NGC~891 observations. However, M51's correlation has much less scatter and tighter 1$\sigma$ confidence intervals -- calculated from the residuals of the observations and the predictions given by the best-fit linear regression coefficients -- compared to that of NGC 891. 

The correlation is also evident in the case of M83 with similar Spearman (0.86) and Pearson (0.92) coefficients. The best fit linear relation deviates away from the 1:1 relationship for the regions of brighter 24~$\mu$m flux density, unlike M51 and NGC~891, and has a gradient and intercept of 0.50$\pm$0.01 and -5.42$\pm$0.03, respectively. The difference in the correlation appears to be driven by the central pixels which are very bright in the 24~$\mu$m band yet have lower \nii \ line emission than we would expect from the other two cases. \citet{bendo2012} warn that the central 8$\arcsec$ of the galaxy is saturated in the 24 $\mu$m map. In the event that this saturation causes some issue in our convolution and rescaling of the 24 $\mu$m map to match the properties of the \nii \ intensity map, we also examine the best fit linear relation when removing these central pixels (grey lines in \hyperref[fig:correlation]{Fig.~\ref*{fig:correlation}g}), finding little variation in slope or intercept of the linear best fit. Finally, NGC~4038/9 displays an albeit weaker correlation (\hyperref[fig:correlation]{Fig.~\ref*{fig:correlation}h}). In \hyperref[tab:summary]{Table~\ref*{tab:summary}}, we summarise the coefficients of the best linear fits.

\subsection{\nii \ 205~$\mu$m line as a SFR tracer}

We suspect the correlation between the \nii \ 205~$\mu$m line emission and the 24~$\mu$m continuum may be originating from the capability of both quantities to trace the star formation activity on spatially-resolved scales. Even if the \nii \ line emission and SFR both correlate with the 24~$\mu$m emission, this does not automatically imply that the \nii \ line should also correlate with the SFR. For example, the SFR could also be dependent on the diffuse UV emission not associated with the warm ISM around young stars. An immediate test of this hypothesis is to investigate the relationship of the \nii \ 205~$\mu$m line emission from each region to the local SFR surface density, $\Sigma_{SFR}$. Although the 24~$\mu$m band is often used as a star formation tracer (see e.g. \citealp{calzetti2005,calzetti2007}), it is important to test any relation between \nii \ and  $\Sigma_{SFR}$ independent of the 24~$\mu$m continuum. We therefore estimate $\Sigma_{SFR}$ with a multi-wavelength composite tracer (see e.g. \citealp*{kennicutt2012}), adopting the SFR calibration of \citet{hao2011} for a linear combination of the FUV flux from Galaxy Evolution Explorer (GALEX, \citealp{martin2005}) observations and mid-IR flux from Wide-field Infrared Survey Explorer (WISE, \citealp{wright2010}) 22 $\mu$m maps (both convolved and rescaled to match the resolution and pixel size of the \nii \ map), in order to correct the UV for dust attenuation independent of the MIPS 24~$\mu$m data. However, we also examined the trends using the SFR based on the 24~$\mu$m data, $\Sigma_{SFR}^{\mathrm{FUV+24}}$, and note, as previously observed by \citet{lee2013}, very little difference between the two tracers.

In the right panels of \hyperref[fig:correlation]{Fig.~\ref*{fig:correlation}}, we compare the \nii \ 205~$\mu$m line emission to the SFR surface density, $\Sigma_{SFR}^{\mathrm{FUV+22}}$. Remarkably, the \nii \ emission correlates with the SFR in a similar fashion as the observed 24~$\mu$m flux density in all five galaxies (i.e. compare the middle and right columns). Again, we find that the relations for the discs of NGC~891 and M51 are similar, yet M83 and NGC 4038/9 have shallower slopes (see \hyperref[tab:summary]{Table~\ref*{tab:summary}}). Interestingly, the former two objects form a relation lying almost parallel to that seen for the integrated galaxy quantities in \citet{zhao2013}, whereas NGC 4038/9 and the central regions of M83 have much higher SFRs than their \nii \ line emission would suggest (see also Fig.~8 of \citealp{wu2014}). 

\begin{figure}
\begin{center}
\includegraphics[width=0.85\columnwidth]{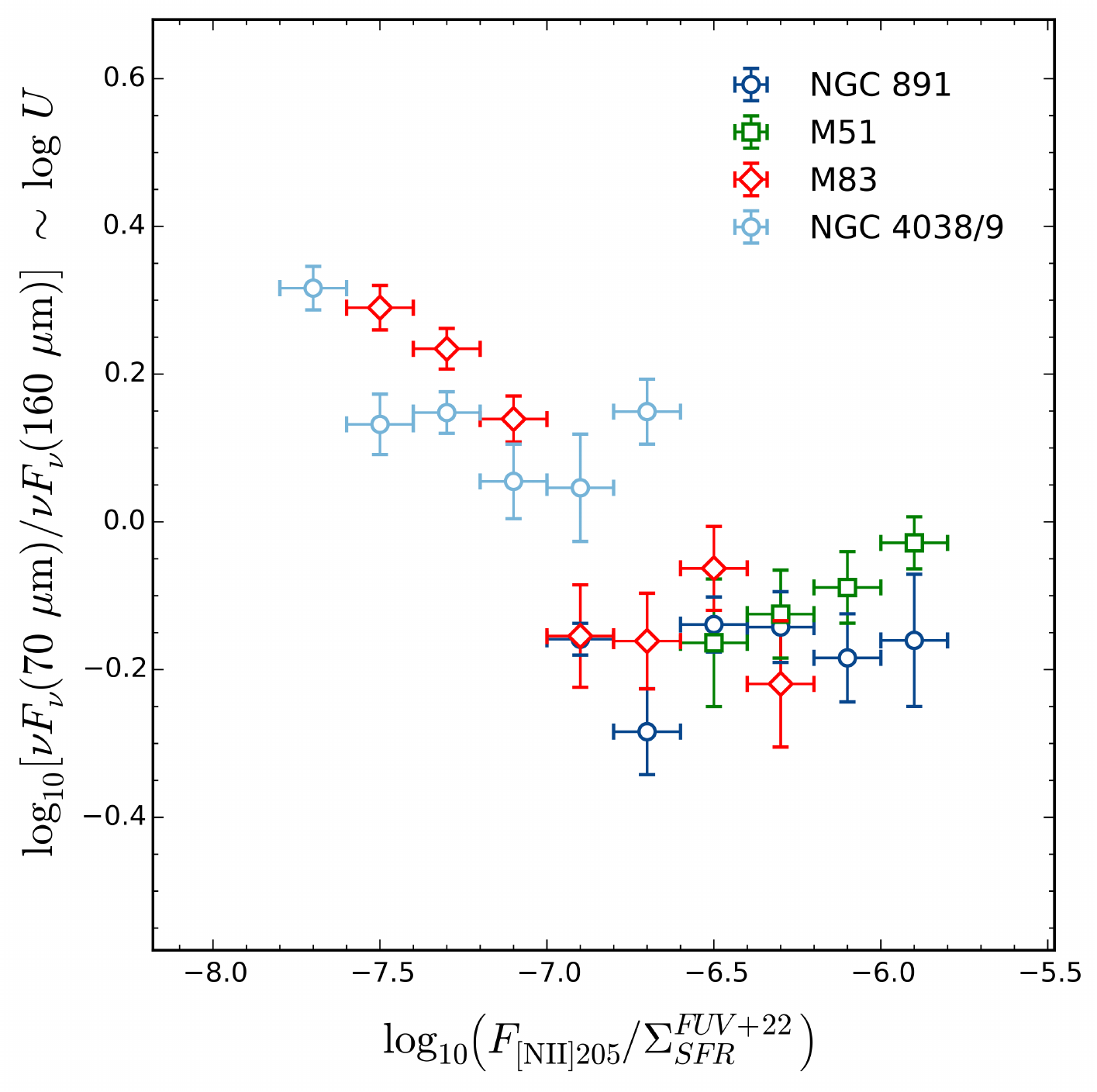}
\end{center}
\vspace{-0.5cm}
\caption{The $\nu F_{\nu}(70~\mu\mathrm{m})/\nu F_{\nu}(160~\mu\mathrm{m})$ FIR colour versus the ratio of the \nii \ emission in W~m$^{-2}$~sr$^{-1}$ to the SFR in M$_{\odot}$~yr$^{-1}$~kpc$^{-2}$ for each galaxy, where points represent the median in bins along the abscissa of width indicated by the errorbars.}\label{fig:ionpot}
\end{figure}

Perhaps the fact that the \nii -- $\Sigma_{SFR}$ correlation differs in the cases of M83 and NGC~4038/9 compared to the other two normal star-forming galaxies is not so surprising, for the following reasons. Firstly, the center of M83 is atypical in that it exhibits a double nucleus in addition to a nuclear starburst (see \citealp{wu2014}, and references therein). The higher SFRs driven by the starburst partly govern the slope of best-fit \nii \ 205 -- $\Sigma_{SFR}$  relation; omitting the central nine pixels covering the nucleus and bar from the linear regression uncovers a best-fit line with a slope similar to that found for the other two galaxies (see the grey lines in \hyperref[fig:correlation]{Fig.~\ref*{fig:correlation}k}). Similar to the nucleus of M83, the Antennae galaxies also exhibit high SFRs, likely from merger-triggered star formation: estimates suggest a stellar mass of $\sim$2.5$\times$10$^8$ M$_{\odot}$ has been formed in $\sim$7$\times$10$^{6}$ years \citep{fischer1996}. Secondly, the SPIRE FTS observations of NGC 891 and M51 cover a greater fraction of the star-forming regions as traced by the 24~$\mu$m emission than those of M83, where nearly 40\% of the pixels cover the inter-arm regions with lower levels of star formation (see \hyperref[fig:correlation]{Fig.~\ref*{fig:correlation}}, left column). We can only speculate that a hypothetical increase of the spatial coverage of the M83 observations to include more of the spiral arms / star-forming disc (and still omitting the nucleus) would yield a \nii \ 205 -- SFR relationship with slope and scatter similar to the other two galaxies.

\section{Discussion}

We have shown that the \nii \ 205 -- 24~$\mu$m emission correlation is present in nearby face-on star-forming discs, indicating the relationship is unlikely to be an artefact arising from line-of-sight projection effects and lending credence to the notion of a physical origin. In addition, a preliminary analysis of VNGS target NGC~2403 yields similar results (Hughes et al. in preparation). At present, the strong correlation between the \nii \ emission and the star formation rate surface density -- traced with two independent measures of the mid-IR continuum -- is suggestive that the recent star formation history is driving this empirical relation.  

The limitations of this analysis are clear. Firstly, as mentioned in \citet{hughes2014a}, despite the high quality of observations of the far-infrared fine-structure lines made possible by the \textit{Herschel Space Observatory}, enabling us to resolve features on sub-kiloparsec scales, one of the main restrictions was the relatively sparse coverage and low spatial resolution of the \nii ~205~$\mu$m line. There is also uncertainty in the SFR calibration; since the FUV emission predominately traces massive ($M > 10-15$~M$_{\odot}$) stars that typically survive for $<$30~Myr, whereas the 22 and 24~$\mu$m emission traces mostly dust in \hii\ star-forming regions, probing lower mass ($\sim$ 4~M$_{\odot}$) stars with lifetimes of a few hundred~Myr on the main sequence, they both probe different time scales of recent star formation (see \citealp*{kennicutt2012}). Different star formation histories for different regions will lead to changes in the relative importance of the FUV and the 22 or 24~$\mu$m emission, and so may therefore produce regional variations in the calibration (e.g. \citealp{boquien2014}). Further complications arise when considering that asymmetric migration of photoionizing stars $<$4~Myr in age from their birth sites in spiral density waves may introduce a spatial offset in UV emission and 24~$\mu$m emission (see Fig. 8 in \citealp{jones2015}), and that stochastically-heated dust associated with \hi \ gas also emits in the 22~$\mu$m band (see e.g. \citealp{eufrasio2014}).

Momentarily casting these issues aside, the fact remains that, if we adopt the correlations found in the cases of NGC~891 and M51, then the higher star formation rates in NGC4038/9 and the center of M83 associated with a nuclear starburst would be under-predicted by the observed \nii \ 205 $\mu$m line flux. Practically, the \nii \ 205 $\mu$m line is not an ideal SF tracer. This raises the question of by what mechanism(s), or under which physical conditions, does the \nii ~205~$\mu$m line cease to be a tracer of star formation and/or the underlying nature of the correlation change. One possible explanation arises when we consider that heating in ionized gas is mainly due to photoionization, and so therefore depends strongly on the critical electron density and the ionization parameter, $U$, defined as the dimensionless ratio of the number density of incident ionizing photons to the number density of hydrogen nuclei. Theoretical models predict that IR colours can act as \textit{approximate} tracers of $U$, whereby $\log U$ increases with the IR colour (see e.g. the $\nu F_{\nu}(60~\mu\mathrm{m})/\nu F_{\nu}(100~\mu\mathrm{m})$ colours in \citealp{abel2009}). 

In \hyperref[fig:ionpot]{Fig.~\ref*{fig:ionpot}}, we plot the $\nu F_{\nu}(70~\mu\mathrm{m})/\nu F_{\nu}(160~\mu\mathrm{m})$ FIR colour\footnote{Homogeneous VNGS \textit{Herschel} PACS and SPIRE photometric maps are available on HeDaM at \burl{http://hedam.lam.fr/VNGS/data.php}.} versus the ratio of the \nii \ emission (in W m$^{-2}$ sr$^{-1}$) to the SFR (in M$_{\odot}$ yr$^{-1}$ kpc$^{-2}$) for each galaxy. The range and mean of the FIR colour is evidently similar for the normal star-forming discs, except for a clear offset in the FIR colour of M83 and NGC~4038/9 that coincides with low \nii \ emission per SFR values. Similar trends emerge when alternatively using the 70/250 and 160/250~$\mu$m FIR colours. Our observations suggest that the ionization parameter is much higher in these regions compared to the rest of the M83 galaxy and the other two discs. Therefore, we expect higher ionization lines (e.g. \niii \ 57~$\mu$m) to play a greater role in the cooling of the gas, which could also partly explain the lower $F_{\mathrm{[NII]}205}/\Sigma_{SFR}$ ratios. It is also possible that dust obscuration becomes more effective in more compact star forming regions (i.e. at increasing FIR colour), such that more ionizing photons go towards dust heating instead of gas heating, leading to a decreasing \nii$/\Sigma_{SFR}$ ratio (see \citealp{diazsantos2013}). Furthermore, we remind the reader that the \citet{zhao2013} \nii \ 205 -- SFR relation was defined based on observations of LIRGS, which are known to have strong ionization parameters (e.g. \citealp{petric2011}), so it may not be so surprising that the M83 nucleus with its high $U$ falls on this relation (as noted by \citealp{wu2014}) in contrast to the behaviour observed in regions with low values of the ionization parameter.

\begin{figure}
\begin{center}
\includegraphics[width=0.85\columnwidth]{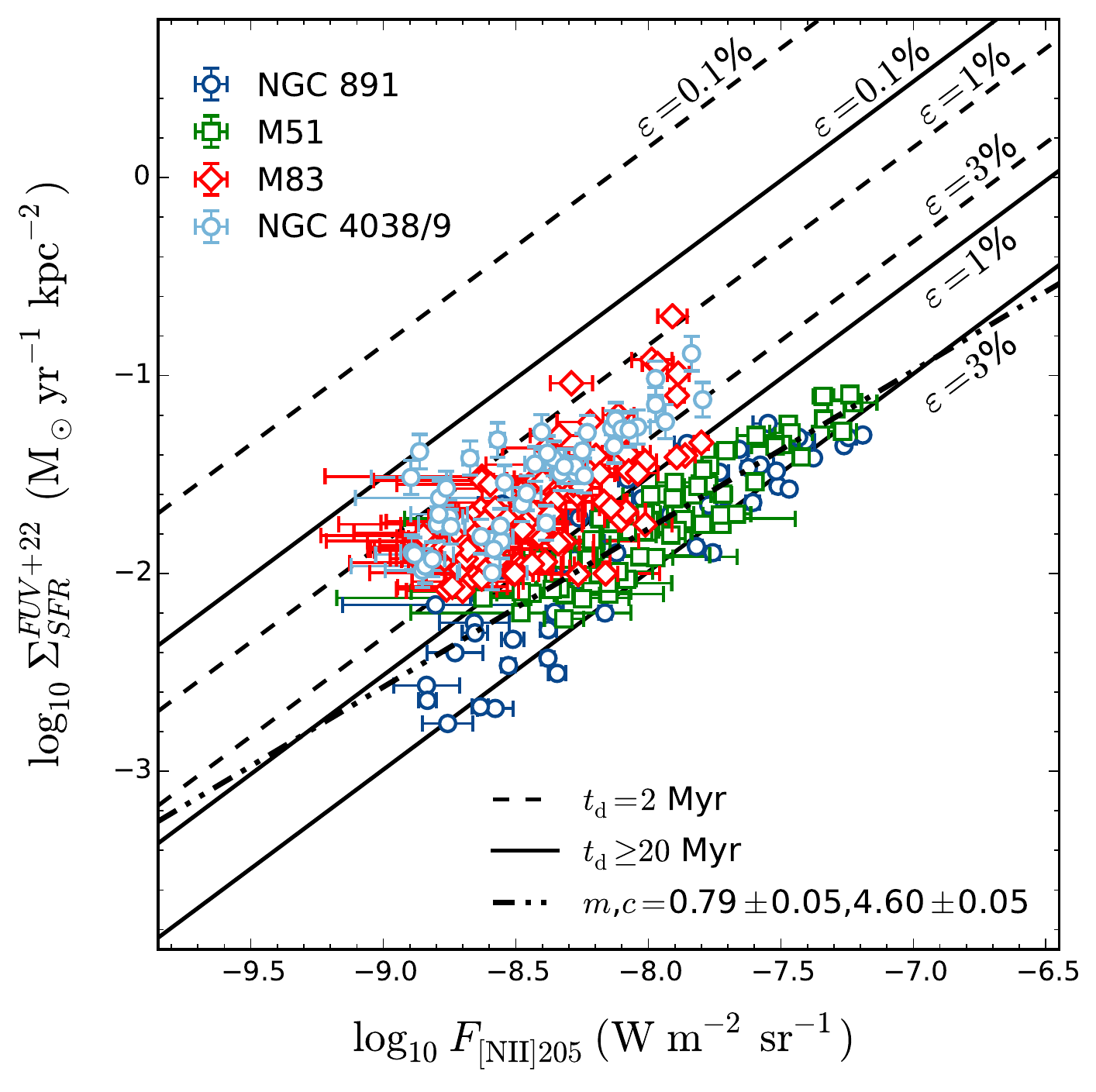}
\end{center}
\vspace{-0.5cm}
\caption{The $\Sigma_{SFR}$ -- $F_{\mathrm{[NII]205}}$ correlation superimposed on the predictions from the \textit{Starburst99} model. For varying values of $\varepsilon$, the dashed lines correspond to a stellar population with a constant SFR and $t_{\mathrm{d}}=2$~Myr, and solid lines correspond to $t_{\mathrm{d}}\geq 20$~Myr. The dash-dotted line is the best-fitting linear relation from the M51 and NGC 891 data.}\label{fig:sfrcal}
\end{figure}

From these results, we attempt to derive a general relation between the \nii ~205~$\mu$m line flux density and star formation rate for main-sequence star-forming galaxies. We first discard regions with very high ($>0.2$, see \hyperref[fig:ionpot]{Fig.~\ref*{fig:ionpot}}) 70/160~$\mu$m FIR colours, which in practice is equivalent to omitting pixels where $\log_{10} F_{24} (\mathrm{W~m}^{-2}~\mathrm{sr}^{-1}) > -5.1$ (c.f. \hyperref[fig:correlation]{Fig.~\ref*{fig:correlation}g} and \hyperref[fig:correlation]{\ref*{fig:correlation}h}), and examine the $\Sigma_{SFR}$--$F_{\mathrm{[NII]}205}$ correlations for all galaxies (see \hyperref[fig:sfrcal]{Fig.~\ref*{fig:sfrcal}}). Simply combining all the data to define an `average' relation is prevented by the aforementioned offset in the individual galaxy relations. To interpret this offset, we use the {\it Starburst99} stellar population synthesis code \citep{leitherer1999} to model the $\Sigma_{SFR}$--$F_{\mathrm{[NII]}205}$ relation following the approach of \citet{herreracamus2014}, where gas heating by FUV photons emitted from star-forming regions relate to gas cooling from the \nii \ line via the photoelectric effect in PAHs and dust grains. In brief, we assume a stellar population with a constant SFR over 100~Myr and solar metallicity\footnote{For \mbox{$Z=0.2Z_{\odot}$}, the $y$-intercepts of the model lines in \hyperref[fig:sfrcal]{Fig.~\ref*{fig:sfrcal}} increase by a factor of 0.01 (i.e. lines shift upwards).}, and adopt the Geneva evolutionary tracks (zero rotation) and Kroupa initial mass function. From the resulting spectrum, given as a function of the duration of the star formation episode, $t_{\mathrm{d}}$, we integrate across the energy range of photons that dominate the dust grain photoelectric heating, specifically $6 < E_{\gamma} < 13.6$~eV. The product of the fraction of these FUV photons responsible for the photoelectric heating of dust grains and the photoelectric heating efficiency of the dust gives the overall heating efficiency, $\varepsilon$, which we vary between 0.1 and 3$\%$ to estimate the amount of gas heating. We further assume that the cooling of the gas is typically dominated by the \cii \ transition and that the \cii /\nii 205 line ratio theoretically varies around $\sim$4 for a range of electron densities (see \citealp{oberst2006}). The \nii\ and FUV flux densities are then related by $F_{\mathrm{[NII]}} \sim 0.25 \ \varepsilon \ F_{\mathrm{FUV}}(t_{\rm d})$ for a given $\Sigma_{SFR}$. 
  
Comparing the model predictions to the observations in \hyperref[fig:sfrcal]{Fig.~\ref*{fig:sfrcal}}, we find the vast majority of regions in NGC 891 and M51 may have been actively star-forming for $t_{\mathrm{d}}\geq 20$~Myr with $\varepsilon$ between 1 and 3\%. Due to the model degeneracy between $t_{\mathrm{d}}$ and $\varepsilon$ (see \citealp{herreracamus2014} for details), M83 and NGC 4038/9 exhibit a stellar population with either a star formation duration of $t_{\mathrm{d}}\geq 20$~Myr and $\varepsilon\lesssim 1$\% or more recent star formation episodes of $t_{\mathrm{d}}=2$~Myr and $\varepsilon$ between 1 and 3\%. This suggests these two galaxies also possess regions with higher SFRs from recently-triggered star formation, likely due to the merger (NGC 4038/9) or central starburst (M83), as we discuss above. We thus combine the NGC 891 and M51 data, representing normal star-forming regions, to define the best-fit relation as 
\begin{eqnarray}\label{eqn:allfit}
\log_{10}\, \Sigma_{SFR} = (0.79\pm 0.05)\,\log_{10}\, F_{\mathrm{[NII]}205} + (4.60 \pm 0.05)
\end{eqnarray}
\noindent and valid in the ranges of $-8.9<\log_{10}~F_{\mathrm{[NII]}205}<-7.5$ and $-2.4<\log_{10}~\Sigma_{SFR}<-1.2$, where units are as stated previously. The scatter, defined as the standard deviation of the residuals between the observed and predicted $\Sigma_{SFR}$, is $\pm$0.10 dex. A similar yet offset relation to Eqn.~\ref{eqn:allfit} is formed by M83 and NGC~4038/9, with a gradient $m$ of $0.69\pm0.05$, $y$-intercept of $4.38\pm0.05$ and a 0.16 dex scatter. We stress, however, all of the caveats and issues we mention above, especially the small sample size, that should be considered when using these relations for estimating the star formation rate.

\section{Conclusions}

In this paper, we aimed to test whether the \nii \ 205 -- 24~$\mu$m emission correlation observed in NGC 891 arises from line-of-sight projection effects in an edge-on disc. Using \textit{Herschel} and \textit{Spitzer} observations of two nearby face-on galaxies, M51 and M83, and the interacting Antennae galaxies NGC 4038 and 4039, we have shown that not only is this empirical relationship also observed in face-on galaxies, but also that the correlation appears to be governed by the star formation rate (SFR). Both the nuclear starburst in M83 and the merger-induced star formation in NGC 4038/9 exhibit less \nii \ emission per unit SFR surface density than the normal star-forming discs. These regions of intense star formation exhibit stronger ionization parameters, as traced by the 70/160 $\mu$m far-infrared colour, that suggest the presence of higher ionization lines that may become more important for gas cooling, thereby reducing the observed \nii ~205~$\mu$m line emission in regions with higher star formation rates. Alternatively, dust obscuration may become more effective in more compact star forming regions, meaning more ionizing photons go towards dust heating instead of gas heating. We used a simple model based on \textit{Starburst99} to predict the $\Sigma_{SFR}$--$F_{\mathrm{[NII]}205}$ relation, finding that regions with more recent star formation in the merging Antennae galaxies NGC 4038/9 and the central starburst in M83 likely cause the offsets in \nii \ emission per unit SFR surface density found between these galaxies and the normal star-forming discs. Finally, we present a general relation between the \nii ~205~$\mu$m line flux density and star formation rate for main-sequence star-forming galaxies.

Clearly, future studies should seek to analyse spatially-resolved observations of other FIR lines (e.g. \nii \ 122, \ci ~370~$\mu$m) including not only the central regions of galaxies, but also the sites of star formation along the spiral arms for a larger sample. Although observations of \nii \ 205~$\mu$m can be followed up over a large redshift range by facilities such as NOEMA and ALMA, which can observe the line emission at $z\sim$0.6 in band-10 and at $z\sim$3.2 in band-7, we caution the reader that care should be taken in the use of the line as a star formation tracer given the apparent dependency of the local \nii \ emission on the ionisation parameter.

\renewcommand{\arraystretch}{1.2}
\begin{table}
 \centering
 \begin{minipage}{\columnwidth}
  \caption{The best-fit relations between the \nii \ 205~$\mu$m line emission and various parameters considered in this work, expressed as $\log_{10} (F_{\mathrm{[NII]}205}/$W m$^{-2}$ sr$^{-1})\,=\,c+m\log_{10}\,x$, with the corresponding 1$\sigma$ errors. We also state the scatter in dex ($\sigma$), Spearman ($\rho_\mathrm{S}$) and Pearson ($\rho_\mathrm{P}$) coefficients. For M83, the limited case excludes the central pixels ($\log_{10} F_{24} (\mathrm{W~m}^{-2}~\mathrm{sr}^{-1}) > -4.4$).\vspace{-0.2in}}\label{tab:summary}
\begin{center}
  \begin{tabular}{l c c c c c}

  \hline
\hline
Galaxy & $m$ & $c$ & $\sigma$ & $\rho_\mathrm{P}$ & $\rho_\mathrm{S}$\\
\hline
\multicolumn{6}{l}{$x\equiv$ $F_{24}$ / W m$^{-2}$ sr$^{-1}$}\\
\hline
NGC 891       & 0.77$\pm$0.01 & -3.31$\pm$0.04 & 0.36 & 0.89  & 0.94\\
M51           & 0.83$\pm$0.02 & -2.90$\pm$0.09 & 0.16 & 0.92  & 0.91\\
M83 (all)     & 0.50$\pm$0.01 & -5.42$\pm$0.03 & 0.37 & 0.92  & 0.86\\
M83 (limited) & 0.58$\pm$0.02 & -4.93$\pm$0.08 & 0.20 & 0.94  & 0.86\\
NGC 4038/9 & 0.40$\pm$0.02 & -6.04$\pm$0.05 & 0.40 & 0.56  & 0.80\\
\hline
\multicolumn{6}{l}{$x\equiv$ $\Sigma_{SFR}^{\mathrm{FUV+22}}$ / M$_{\odot}$ yr$^{-1}$ kpc$^{-2}$ }\\
\hline
NGC 891       & 0.94$\pm$0.02 & -6.31$\pm$0.04 & 0.22 & 0.87  & 0.94\\
M51           & 1.07$\pm$0.02 & -6.09$\pm$0.04 & 0.14 & 0.94  & 0.92\\
M83 (all)     & 0.71$\pm$0.01 & -7.27$\pm$0.01 & 0.28 & 0.87 &  0.93\\
M83 (limited) & 1.04$\pm$0.02 & -6.71$\pm$0.04 & 0.19 & 0.78  & 0.74\\
NGC 4038/9 & 0.84$\pm$0.03 & -7.22$\pm$0.05 & 0.22 & 0.65  & 0.79\\
\hline
\multicolumn{6}{l}{$x\equiv$ $\Sigma_{SFR}^{\mathrm{FUV+24}}$ / M$_{\odot}$ yr$^{-1}$ kpc$^{-2}$}\\
\hline
NGC 891       & 0.72$\pm$0.01 & -6.59$\pm$0.02 & 0.35 & 0.87  & 0.91\\
M51           & 0.90$\pm$0.02 & -6.29$\pm$0.03 & 0.14 & 0.92  & 0.90\\
M83 (all)     & 0.66$\pm$0.01 & -7.19$\pm$0.01 & 0.33 & 0.87  & 0.92\\
M83 (limited) & 0.86$\pm$0.01 & -6.79$\pm$0.02 & 0.16 & 0.81  & 0.87\\
NGC 4038/9 & 0.52$\pm$0.02 & -7.52$\pm$0.03 & 0.39 & 0.88  & 0.84\\
\hline
\end{tabular}
\end{center}
\end{minipage}
\end{table}

\section*{Acknowledgments}
We thank the anonymous referee for the useful comments and suggestions that helped to improve this paper. TMH and EI acknowledge CONICYT/ALMA funding Program in Astronomy/PCI Project N$^\circ$:31140020. TMH also acknowledges the financial support from the Belgian Science Policy Office (BELSPO) in the frame of the PRODEX project C90370 (Herschel-PACS Guaranteed Time and Open Time Programs: Science Exploitation). MB and IDL acknowledge the financial support of the Flemish Fund for Scientific Research (FWO-Vlaanderen). SPIRE has been developed by a consortium of institutes led by Cardiff University (UK) and including Univ. Lethbridge (Canada); NAOC (China); CEA, LAM (France); IFSI, Univ. Padua (Italy); IAC (Spain); Stockholm Observatory (Sweden); Imperial College London, RAL, UCL-MSSL, UKATC, Univ. Sussex (UK); and Caltech, JPL, NHSC, Univ. Colorado (USA). This development has been supported by national funding agencies: CSA (Canada); NAOC (China); CEA, CNES, CNRS (France); ASI (Italy); MCINN (Spain); SNSB (Sweden); STFC, UKSA (UK); and NASA (USA). This research made use of APLpy, an open-source plotting package for Python hosted at \url{http://aplpy.github.com}.

\bibliography{aa27644-16}

\end{document}